\documentclass[%
 aip,
 amsmath,amssymb,
 reprint,%
]{revtex4-1}
\usepackage[english]{babel}

\usepackage{graphicx}% Include figure files
\usepackage{dcolumn}% Align table columns on decimal point
\usepackage{bm}% bold math
\usepackage[utf8]{inputenc}
\usepackage[T1]{fontenc}
\usepackage{mathptmx}
\usepackage{newunicodechar}
\newunicodechar{−}{\ensuremath{-}}
\usepackage{etoolbox}
\usepackage{subcaption}
\usepackage{chemformula}
\usepackage{cleveref}
\usepackage{siunitx}
\usepackage{booktabs}
\usepackage{microtype}

\makeatletter
\def\@email#1#2{%
 \endgroup
 \patchcmd{\titleblock@produce}
  {\frontmatter@RRAPformat}
  {\frontmatter@RRAPformat{\produce@RRAP{*#1\href{mailto:#2}{#2}}}\frontmatter@RRAPformat}
  {}{}
}%
\makeatother
\begin{document}

\preprint{AIP/123-QED}

\title[Coupling between ferroelectric distortions and excitonic properties in PbTiO$_3$]{Coupling between ferroelectric distortions and excitonic properties in PbTiO$_3$}
% Force line breaks with \\
\author{Pietro Pacchioni}
\affiliation{Dipartimento di Fisica e Astronomia, Università di Bologna, $40127$ Bologna, Italy}
\affiliation{Institute of Physics, École Polytechnique Fédérale de Lausanne (EPFL), 1015 Lausanne, Switzerland}

\author{Lorenzo Varrassi}
\affiliation{Dipartimento di Fisica e Astronomia, Università di Bologna, $40127$ Bologna, Italy}
\affiliation{CINECA National Supercomputing Center, Casalecchio di Reno, I-40033 Bologna, Italy}

\author{Cesare Franchini}
\affiliation{University of Vienna, Faculty of Physics, Center for Computational Materials Science, Vienna, Austria}
\affiliation{Dipartimento di Fisica e Astronomia, Università di Bologna, $40127$ Bologna, Italy}

\date{\today}% It is always \today, today,

\begin{abstract}
PbTiO$_3$ is a ferroelectric perovskite semiconductor with favourable electronic and optical properties, making it suitable for a wide range of applications, including photo-catalysis and (opto)electronic devices.
Despite its relevance, an accurate \emph{ab-initio} description of the optical absorption spectrum and of the impact of ferroelectric distortion on the excitonic properties is still lacking. 
We combine $G_0W_0$ and Bethe–Salpeter Equation calculations to investigate the electronic and optical properties of PbTiO$_3$, tracking the evolution of its excitonic spectrum along the transition from the cubic paraelectric to the tetragonal ferroelectric phase.
As the polar distortion increases, the first absorption peak of the cubic phase splits into two distinct features due to symmetry breaking, which partially lifts the degeneracy of the underlying excitonic state. 
Crucially, the distortion further introduces an in-plane/out-of-plane anisotropy in the spectra and controls the energy separation between the resulting excitonic branches.
These findings highlight the potential for tuning the optical absorption properties of PbTiO$_3$ via the application of an external electric field.
\end{abstract}

\maketitle

\section{Introduction}
Transition‑metal‑oxide (TMO) perovskites present a rich interplay of charge, spin, orbital and lattice degrees of freedom underpinning phenomena ranging from colossal magneto‑resistance \cite{von_helmolt_giant_1993, salamon_physics_2001} to multiferroicity \cite{wang_multiferroicity_2009} and superconductivity \cite{bednorz_possible_1986, tokura_correlated-electron_2003}, which have led to their employment in a variety of applications.

Among ferroelectric TMOs, \ch{PbTiO3} (PTO) stands out thanks to its large spontaneous polarization \cite{abrahams_atomic_1968, haun_thermodynamic_1987}, sizeable piezo‑ \cite{tawfik_high_2018, zhang_size-dependent_2008} and electro‑optic coefficients \cite{fontana_electrooptical_1995, veithen_first-principles_2004} and the possibility of switching its tetragonal distortion with modest electric fields \cite{nishino_evolution_2020, ma_electric_2014}. These attributes make PTO a prototype platform for tunable opto‑electronic devices, non-linear optical systems \cite{fork_application_1996} and solar cells applications \cite{wang_pbtio3_2019}.

Despite growing interest, first-principles descriptions of the optical spectra of PbTiO$_3$ remain limited. Only a few studies report its absorption spectrum \cite{Yaseen_Ambreen_Mehmood_Iqbal_Iqbal_Alshahrani_Noreen_Laref_2021, Hosseini_Movlarooy_Kompany_2005, Bendaoudi_Ouahrani_Daouli_Rerbal_Boufatah_Morales-García_Franco_Bedrane_Badawi_Errandonea_2023}, with most relying on the independent-particle approximation or semi-empirical Hubbard corrections. These approaches neglect electron–hole interactions, which are known to be crucial for accurately reproducing experimental spectra \cite{sponza_role_2013, Liu_Kim_Chen_Sarma_Kresse_Franchini_2018, Begum_Gruner_Pentcheva_2019}. Moreover, only a limited number of studies incorporate a quasiparticle GW framework \cite{Bendaoudi_Ouahrani_Daouli_Rerbal_Boufatah_Morales-García_Franco_Bedrane_Badawi_Errandonea_2023, gou_post_2011}, despite its status as the state-of-the-art approach for modeling (ferroelectric) perovskites \cite{klimes_predictive_2014, He2014, ergonenc_converged_2018}.

Furthermore, the cubic and tetragonal phases have traditionally been studied as distinct, isolated structures, leaving open the question of how the electronic, optical, and excitonic properties evolve continuously with the gradual onset of ferroelectric distortion. Such intermediate polar states are experimentally accessible through the application of an electric field \cite{lim_dynamics_2021, zhai_ferroelectric_2004}, and may enable electric-field control of the light absorption spectrum.

An accurate description of the electronic and optical properties of materials requires both quasiparticle (QP) corrections beyond density functional theory (DFT) and a proper treatment of excitonic effects. Single-shot $\mathrm{G_0W_0}$ calculations have proven predictive for perovskites \cite{varrassi_optical_2021, PhysRevMaterials.3.103801, He2014, Liu_Kim_Chen_Sarma_Kresse_Franchini_2018, ergonenc_converged_2018, Varrassi2025}, while solving the Bethe–Salpeter equation (BSE) enables the inclusion of electron–hole interactions, which are essential for accurately computing exciton binding energies and oscillator strengths \cite{albrecht_ab_1998, onida_electronic_2002}. However, a fully self-consistent GW+BSE workflow on the dense $\bm{k}$-point meshes required for convergence of excitonic energies remains computationally prohibitive \cite{albrecht_ab_1998}. A practical alternative is the model-BSE (mBSE) approach, where the screened Coulomb interaction is replaced by a physically motivated analytical model \cite{bechstedt_efficient_1992, fuchs_quasiparticle_2007, bokdam_role_2016}. Benchmark studies have demonstrated that mBSE can satisfactorily reproduce full GW+BSE spectra for TMOs while reducing the computational cost by orders of magnitude \cite{varrassi_optical_2021,Varrassi2024}.

In this work, we address the aforementioned gaps by performing a systematic $\mathrm{G_0W_0}$ + mBSE investigation of PbTiO$_3$ along a continuous ferroelectric distortion pathway, connecting the paraelectric cubic structure to the fully relaxed tetragonal phase. For seven intermediate configurations, we obtain orbital-projected band structures with converged $\mathrm{G_0W_0}$ parameters, solve the mBSE on dense $\bm{k}$-point grids to extract the optical absorption spectra, and trace the evolution of the fundamental, direct, and optical gaps, as well as the first bright and dark excitons. Our results reveal how the breaking of cubic symmetry lifts the triple degeneracy of the lowest exciton, leading to a polarization-dependent splitting of the first absorption peak.

This splitting of the first cubic absorption peak into two distinct features is correlated with the emergence of anisotropic exciton wavefunctions: one localized along the $\Gamma$–Z direction in $\bm{k}$-space, and the other pinned at $\Gamma$. This behaviour provides a microscopic mechanism for tailoring optical selection rules through controlled ferroelectric displacements.

\section{Methods}\label{sec:methods}

All calculations were performed using the Vienna ab initio Simulation Package (\textsc{vasp}) \cite{kresse_ab_1993, kresse_efficient_1996} with the projector-augmented wave method (PAW) \cite{blochl_projector_1994}. The GW versions of the potentials were used. Norm conserving (NC) potentials including the semicore states were used for the titanium atom to ensure accurate QP energies \cite{klimes_predictive_2014, ergonenc_converged_2018,Varrassi2025}.

Ionic relaxation was performed for both the cubic and tetragonal structures of PTO at the DFT level, in the generalized gradient approximation (GGA) with Perdew, Burke and Ernzerhof (PBE) functionals \cite{perdew_generalized_1997}. PBE is known to overestimate the tetragonal distortion of TMO perovskites \cite{zhang_comparative_2017,PhysRevMaterials.3.103801}, so the structural parameters were interpolated between the relaxed cubic and tetragonal structures, producing seven intermediate structures. A ferroelectric structure closely matching experiment (see \Cref{tab:structure-data}) was selected as representative of this phase of PTO in the absence of electric field. Details about the structures with higher tetragonal distortions are discussed in the Supplementary Materials (SM). 

\begin{table}[h]
    \centering
    \begin{tabular}{c c c c c}
    \toprule
         & $a$ [\unit{\angstrom}] & $c/a$ & $\Omega$ [\unit{\angstrom^3}] & $\Delta_{Ti}$ \\
    \midrule
       Interpolated & 3.926 & 1.069 & 64.66 & 0.034 \\
       Experiment & \numrange[range-phrase=~--~]{3.880}{3.904} & \numrange[range-phrase=~--~]{1.064}{1.071} & \numrange[range-phrase=~--~]{62.56}{63.31} & \num{0.038} \\
    \bottomrule
    \end{tabular}
    \caption{Lattice constant $a$, tetragonal distortion $c/a$, cell volume $\Omega$ and titanium ion displacement $\Delta_{Ti}$ for the selected interpolated tetragonal structure (Interpolated) compared to experiment. Experimental ranges from refs. \cite{yoshiasa_high-temperature_2016, nelmes_crystal_1985, shirane_x-ray_1956, mabud_lattice_1979}}.
    \label{tab:structure-data}
\end{table}

For all structures, the absorption spectra were determined following a procedure similar to that outlined in refs. \cite{varrassi_optical_2021, ergonenc_converged_2018}:
\begin{enumerate}
    \item An initial self-consistent DFT step is performed;
    \item A single shot $G_0W_0$ calculation is performed by constructing the one particle Green's function from the single electron energies and orbitals determined in the previous step, while the screened Coulomb interaction is computed using the Random Phase Approximation (RPA) \cite{shishkin_implementation_2006, del_sole_gwensuremathgamma_1994}. The convergence procedure for the QP energies described in \cite{ergonenc_converged_2018} was employed, leading to a requirement of \num{3200} virtual bands and a plane-wave energy cut-off of \qty{900}{\electronvolt} for convergence to about \qty{0.01}{\electronvolt}. The screened Coulomb interaction and the self-energy were represented with 100 frequency points.
    \item The optical absorption spectra are calculated by solving the BSE using the mBSE approach in the Tamm-Damcoff approximation, starting from the $G_0W_0$ data. %\textbf{How Many frequency points were employed to represent the screened potential and the self-energy?}
\end{enumerate}
Convergence of the $\bm{k}$-grid is critical when solving the BSE \cite{albrecht_albrecht_1999, fuchs_efficient_2008, kammerlander_speeding_2012}. To ensure accurate results, multiple calculations of the absorption spectra of the cubic and tetragonal structures were performed with increasingly dense $\bm{k}$-grids. No new peaks were verified to appear under \qty{5}{\electronvolt} when employing a uniform, $\Gamma$-centred $16\times16\times16$ grid. Convergence was also verified with respect to the number of occupied and unoccupied orbitals included in the calculation, which were set to 6 and 8 respectively. 

The mBSE approximation replaces the RPA dielectric function $\epsilon^{-1}(\bm{k})$ with an analytical model \cite{bokdam_role_2016} and approximates the QP energies by applying a scissor operator (rigid shift) to the DFT results, aligning the direct band gap with that obtained from $\mathrm{G_0W_0}$ calculations. This method has been successfully applied to other TMO perovskites \cite{Begum_Gruner_Pentcheva_2019, varrassi_optical_2021}. Its applicability to the present case was validated by comparison with the $k$-averaging \cite{Liu_Kim_Chen_Sarma_Kresse_Franchini_2018, sander_beyond_2015} and QP interpolation \cite{kammerlander_speeding_2012} approaches (see Supplementary Material).

\section{Results and discussion}

\subsection{GW Electronic Properties}
\begin{figure*}
    \centering
    \includegraphics[width=\linewidth]{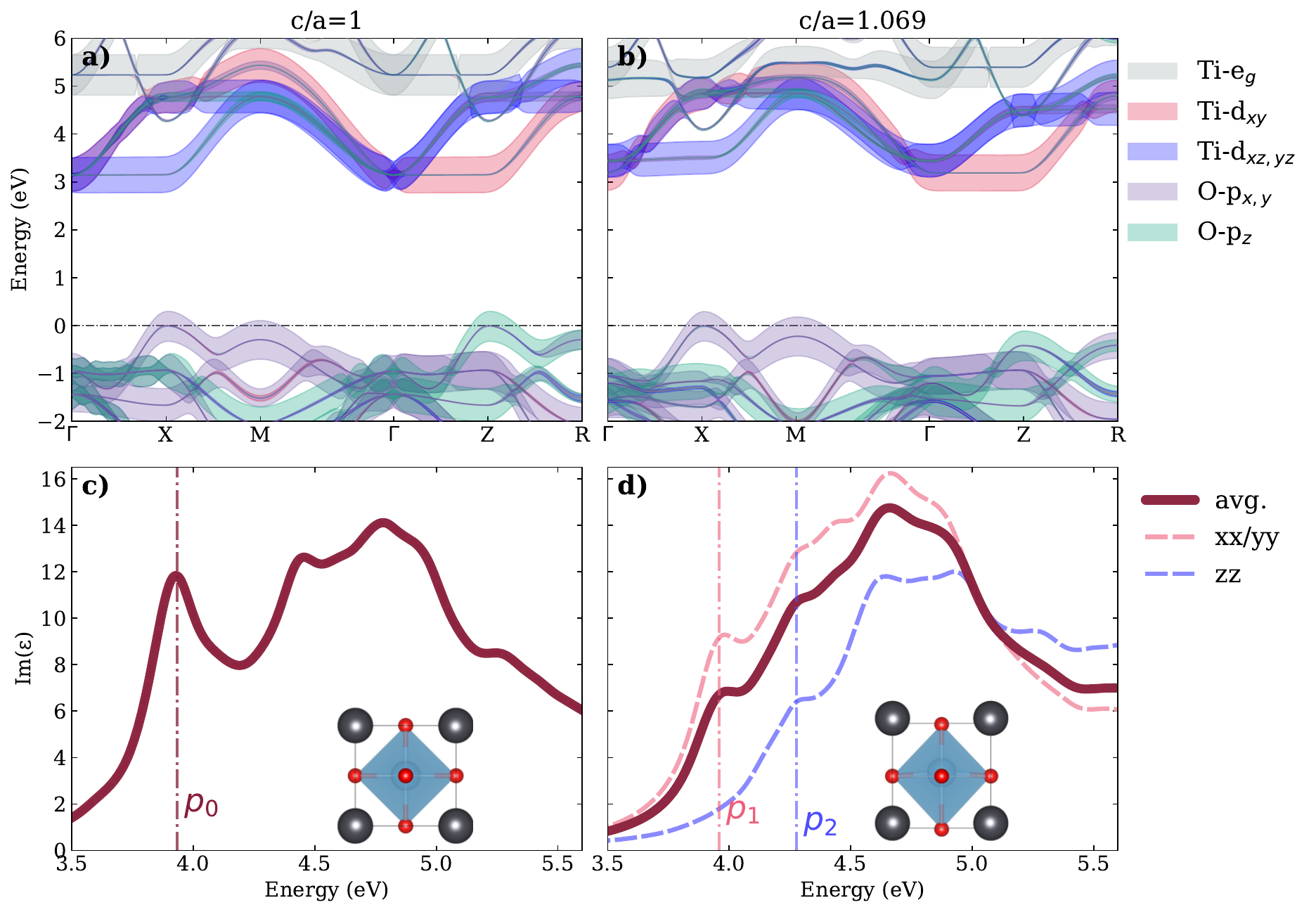}
    \caption{\textbf{a), b)} panels: band structures of the cubic and tetragonal phases of PTO projected on the orbitals of interest. The Fermi energy $E_F$ is set to zero, while the arrows indicate the optical transitions and are coloured coded according to the spectral features they contribute to. \textbf{c), d)} panels: Corresponding mBSE  optical absorption spectra. For the tetragonal phase, the $xx/yy$ and $zz$ components of the dielectric tensor are reported separately, along with the isotropic spectrum, calculated as an average among the components. The dashed vertical lines are located at the energies of the analysed peaks. Insets: Crystalline structure of the two phases of PTO. Pb ions in black, O in red, Ti in light blue.}
    \label{fig:main-fig}
\end{figure*}
The calculated direct and indirect band gaps for both the cubic and tetragonal phases of PbTiO$_3$ computed at $\mathrm{G_0W_0}$ level are reported in \Cref{tab:bandgap-reference}. The complete band structures are shown in Fig.~\ref{fig:main-fig}. As expected from previous studies \cite{jain_high-throughput_2011, zhang_comparative_2017}, DFT severely underestimates the band-gap values, while the $\mathrm{G_0W_0}$ results, which fall within the ranges reported in earlier works \cite{gou_post_2011, Bendaoudi_Ouahrani_Daouli_Rerbal_Boufatah_Morales-García_Franco_Bedrane_Badawi_Errandonea_2023}, provide a much more accurate description of the experimental values \cite{zametin_absorption_1984, peng_optical_1991, lemziouka_structural_2021}. The indirect band gap is located along the X–Z direction in $\bm{k}$-space, while the direct gap occurs at the X point.
\begin{table}[h]
        \centering
        \begin{tabular}{c c c c c}
        \toprule
          & \multicolumn{2}{c}{This work} & \multicolumn{2}{c}{Literature} \\
          & Indirect & Direct & Indirect & Direct \\
        \midrule
        PBE & 1.94 & 2.91 & 1.81\cite{jain_high-throughput_2011}/1.88\cite{zhang_comparative_2017} & 2.95\cite{jain_high-throughput_2011} \\
        GW & 3.20 & 3.49 & 3.90\cite{gou_post_2011}/2.95\cite{Bendaoudi_Ouahrani_Daouli_Rerbal_Boufatah_Morales-García_Franco_Bedrane_Badawi_Errandonea_2023} & - \\
        Exp. & & & & 3.60\cite{zametin_absorption_1984}/3.45\cite{peng_optical_1991}/3.10\cite{lemziouka_structural_2021} \\
       \bottomrule
       \end{tabular}
       \caption{Calculated band-gaps, in \unit{\electronvolt}, for the ferroelectric phase of \ch{PbTiO3}, using both DFT with PBE pseudopotentials and the $G_0W_0$ approximation. Computational and experimental results from the literature are also reported.}
       \label{tab:bandgap-reference}
\end{table}

The band-gaps were also calculated in the $G_0W_0$ approximation for the structures of intermediate distortion, and are tabulated in \Cref{tab:electronic-properties}: while their values increase monotonically with distortion, their positions in $\bm{k}$-space do not vary (see SM for results on over-distorted structures).
\begin{table*}
        \centering
        \begin{tabular}{c c c c c c c c c}
            \toprule
            Distortion & $a$ [\unit{\angstrom}] & $c/a$ & $E_g^{ind}$ [\unit{\electronvolt}] & $E_g^{dir}$ [\unit{\electronvolt}] & $E_g^{opt}$ [\unit{\electronvolt}] & $E_{xb}$ [\unit{\electronvolt}] & $E_{xb}^{(nd)}$ [\unit{\electronvolt}] & $\varepsilon_\infty^{-1}$\\
            \midrule
            \textbf{0\%} & 3.96 & 1 & 3.14 & 3.15 & 3.056 & 0.095 & 0.083 & 0.131 \\
            40\% & 3.95 & 1.027 & 3.14 & 3.19 & 3.097 & 0.098 & 0.086 & 0.132 \\
            60\% & 3.94 & 1.041 & 3.15 & 3.26 & 3.174 & 0.102 & 0.088 & 0.134 \\
            80\% & 3.93 & 1.055 & 3.18 & 3.36 & 3.280 & 0.107 & 0.090 & 0.136 \\
            \textbf{100\%} & 3.93 & 1.069 & 3.20 & 3.49 & 3.403 & 0.118 & 0.092 & 0.139 \\
            \bottomrule
        \end{tabular}
        \caption{Lattice constant $a$, tetragonal distortion ratio $c/a$, indirect band gap $E_g^{\mathrm{ind}}$, direct band gap $E_g^{\mathrm{dir}}$, optical gap $E_g^{\mathrm{opt}}$, exciton binding energy $E_{xb}$, first bright (non-dark) exciton binding energy $E_{xb}^{(\mathrm{nd})}$, and inverse dielectric screening $\varepsilon_\infty^{-1}$ for the interpolated structures, as obtained from $\mathrm{G_0W_0}$ and BSE calculations. The 100\% and 0\% data correspond to the ferroelectric (reference interpolated structure, see Tab.~\ref{tab:structure-data}) and paraelectric structures, respectively, as described in the text.
        }
        \label{tab:electronic-properties}
\end{table*}

The first row of \Cref{fig:main-fig} shows the band structures for both the paraelectric and ferroelectric phases of PbTiO$_3$, obtained using the DFT+scissor approximation and projected onto the relevant orbitals ($p$ orbitals of the oxygen ions and $t_{2g}$, $e_g$ orbitals of the titanium ion). These results were validated by comparison with band structures obtained using $\mathrm{G_0W_0}$ calculations combined with Wannier interpolation~\cite{MOSTOFI2014, Franchini2012} (see SM).

\begin{figure*}[t]
    \centering
    \includegraphics[width=\linewidth]{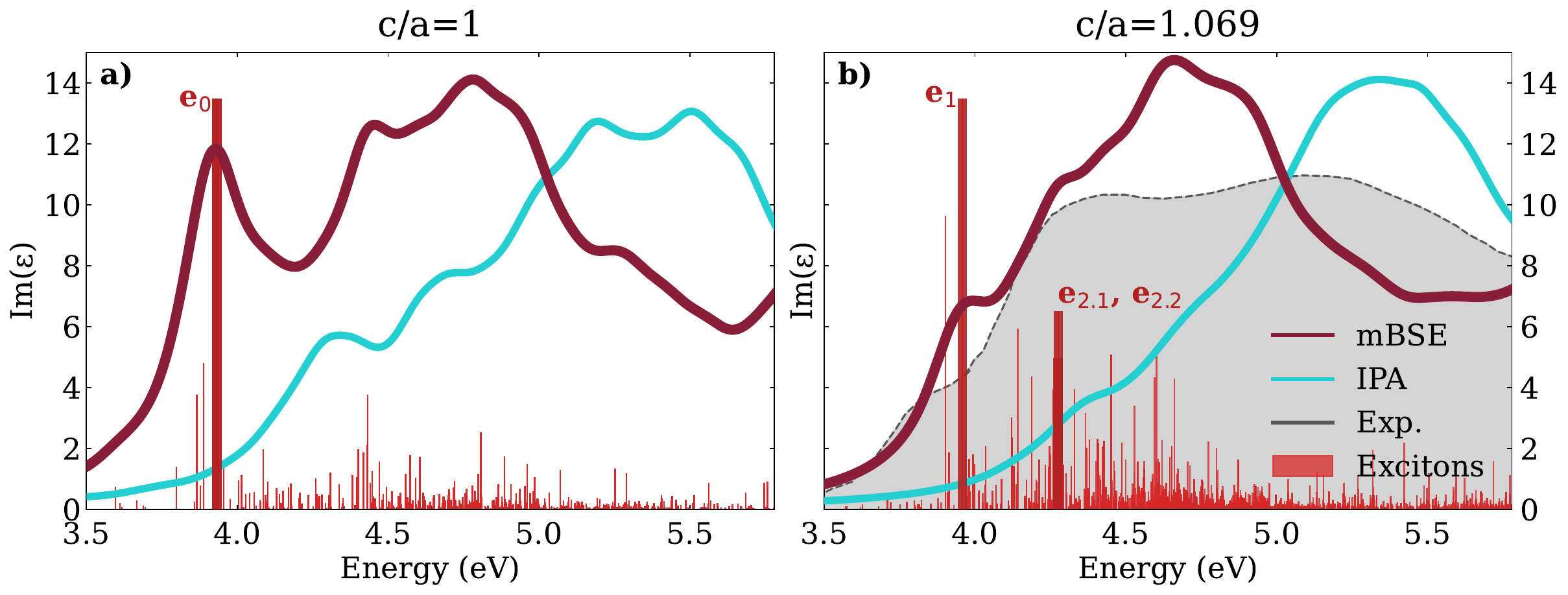}
    \caption{Isotropic optical absorption spectra for the paraelectric and ferroelectric structures obtained in the model-BSE and IPA (Independent Particle) approximations, together with the transition strengths of the main excitons. The excitons contributing to the peaks of interest are indicated by vertical (orange) lines and labelled ($e$). 
    }
    \label{fig:ipa-mbse}
\end{figure*}

As previously reported in the literature \cite{derkaoui_overview_2023, Yaseen_Ambreen_Mehmood_Iqbal_Iqbal_Alshahrani_Noreen_Laref_2021}, the valence bands are predominantly composed of O-$p$ states, while the conduction bands near the band gap are primarily determined by Ti-$d$ states. The latter can be further divided into two contributions: the $t_{2g}$ orbitals ($d_{xy}$, $d_{xz}$, and $d_{yz}$), located at lower energies, and the $e_g$ orbitals ($d_{z^2}$ and $d_{x^2-y^2}$), which appear at higher energies. In both the paraelectric and ferroelectric phases, little mixing occurs between the $t_{2g}$ and $e_g$ states \cite{ergonenc_converged_2018, varrassi_optical_2021, Begum_Gruner_Pentcheva_2019}.
\newline  For the paraelectric phase, the Ti-$t_{2g}$ manifold is threefold degenerate at $\Gamma$ and contributes to the conduction-band minimum (CBM). 
In the ferroelectric tetragonal distortion, the Ti off-centering enhances Ti–O $p$–$d$ hybridization, which plays a key role in stabilizing the ferroelectric phase, and alters the local crystal field, lifting the cubic $t_{2g}$ degeneracy at $\Gamma$.
As a result, the $d_{xy}$ singlet lies at lower energies than the ($d_{xz}$, $d_{yz}$) doublet at $\Gamma$ (~\Cref{fig:main-fig}) and the magnitude of the energy splitting between the Ti-$d_{xz}$/$d_{yz}$ and $d_{xy}$ states increases with increasing distortion amplitude.
\newline Beyond the $\Gamma$ point, the lowest conduction band along $\Gamma$-Z retains its predominant $d_{xy}$ character and remains almost non-dispersive. In contrast, the corresponding band along $\Gamma$-X shifts to higher energies under the tetragonal distortion, consistent with the lifting of the $t_{2g}$ degeneracy at $\Gamma$.

\subsection{Absorption spectra}\label{sec:absorption-spectra}
\begin{figure}[b]
    \centering
    \includegraphics[width=\linewidth]{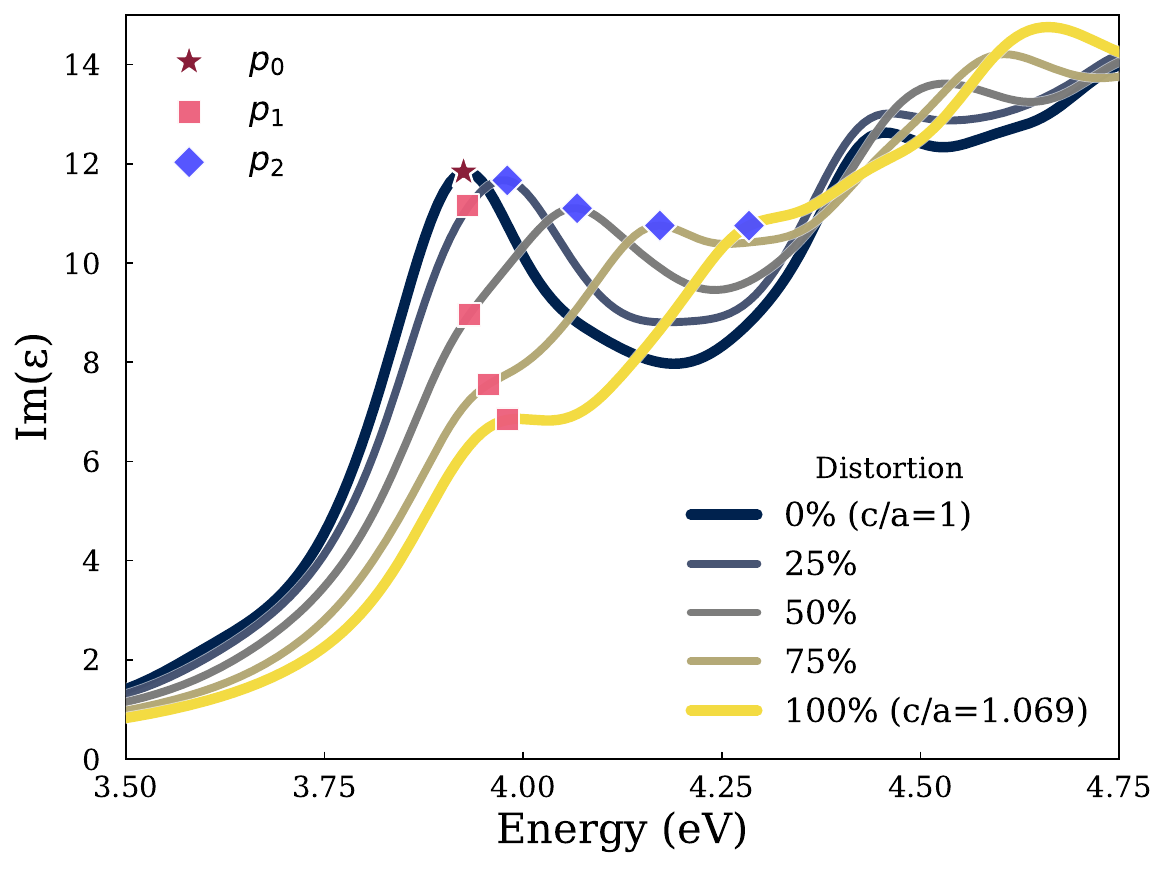}
    \caption{Isotropic optical absorption spectra for intermediate structures along the distortion from the paraelectric to the ferroelectric phase, obtained within the mBSE approximation.
    The marked peak positions highlight how the cubic $p_0$ feature progressively splits into the $p_1$ and $p_2$ branches as the polar distortion increases.
    }
    \label{fig:structures-spectra}
\end{figure}
\begin{figure*}[t]
    \centering
    \includegraphics[width=\linewidth]{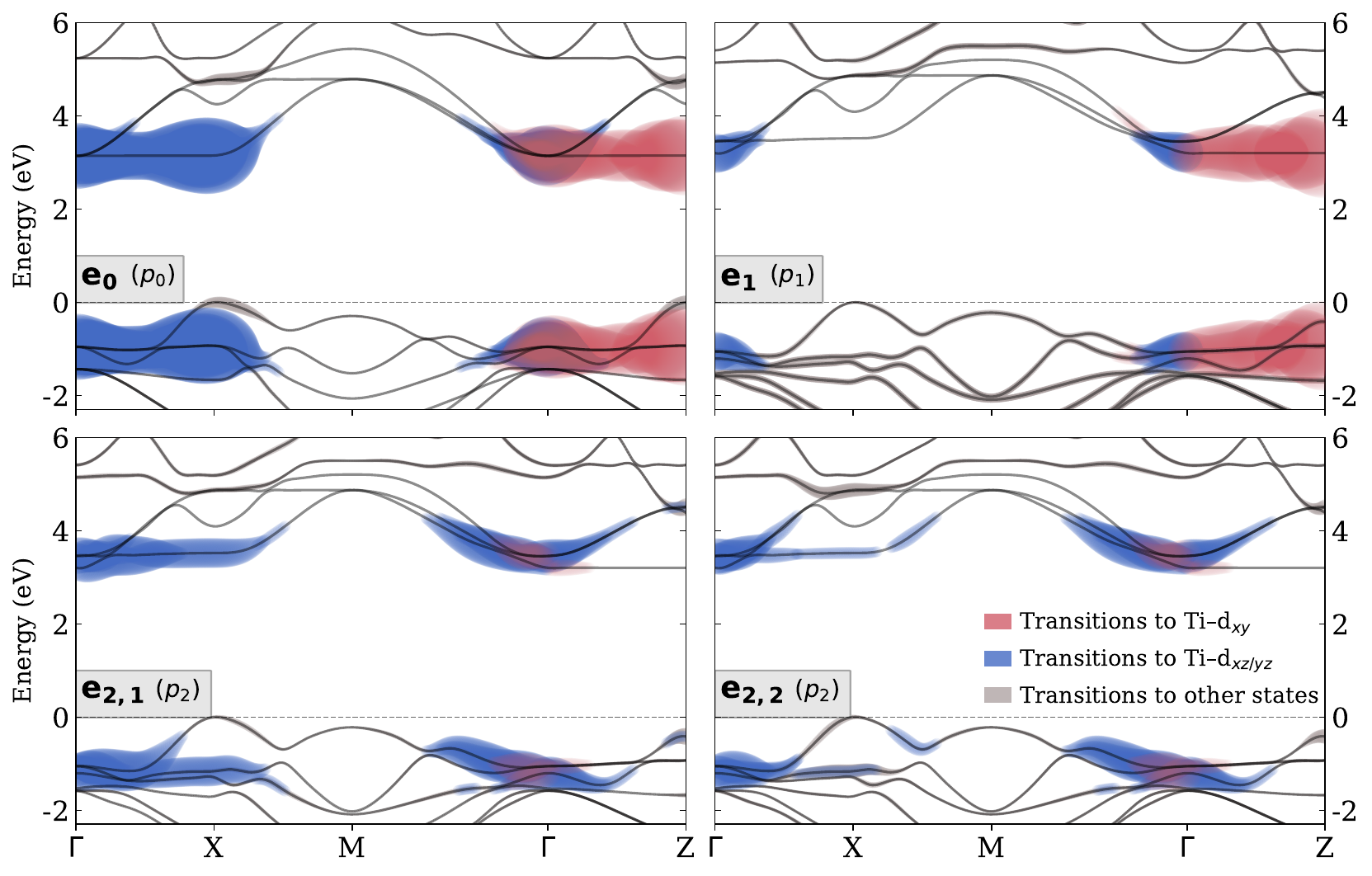}
  \caption{Fatband plots of the excitons from \Cref{fig:ipa-mbse}: exciton $e_0$ from the paraelectric structure splits into three excitons with different characters in the ferroelectric phase. The black dots outline the band structure, while the coloured circles indicate the amplitude of the BSE eigenvector at each point. Different colours refer to different valence-conduction band pairs.} 
\label{fig:fatband}
\end{figure*}

Table \ref{tab:electronic-properties} summarizes the tetragonal distortion, band gaps, and key optical properties for all considered structures, including the optical gap $E_g^{\mathrm{opt}}$, the binding energies $E_{xb}$ of the first dark and non-dark (i.e., with nonzero transition strength) excitons as obtained from BSE calculations, and the static inverse dielectric screening $\epsilon_{\infty}^{-1}$. All these quantities are found to increase monotonically with increasing structural distortion. Moreover, the optical gap remains consistently lower than the $\mathrm{G_0W_0}$-calculated direct gap, confirming the presence of bound excitons.
The second row of \Cref{fig:main-fig} reports the absorption spectra (imaginary part of the dielectric function) for the cubic and tetragonal phases of PbTiO$_3$, calculated within the mBSE approximation.
The spectra are shown both as an average over all directions and with the components along the out-of-plane($zz$) and in-plane($xx/yy$) directions separated.
In the cubic paraelectric phase, the isotropic spectrum presents a first prominent peak, labelled $p_0$, at \qty{3.93}{\electronvolt}. 
A feature at similar transition energies is also identifiable for the ferroelectric phase (labelled $p_1$), although with noticeably reduced oscillator strength.
Crucially, this peak is visible only in the $xx/yy$ component and is entirely absent in the out-of-plane response, which shows no trace of the corresponding excitation. Instead, the $zz$ component develops a distinct shoulder around $\sim 4.3$ eV, labelled $p_2$, which is not present in the cubic structure's spectra.
\newline The importance of electron–hole interactions is highlighted in Figure \ref{fig:ipa-mbse}.
For the paraelectric structure in particular, excitonic effects induce a marked redshift and a substantial enhancement of the $p_0$ peak, consistent with trends reported for other cubic transition-metal-oxide perovskites~\cite{sponza_role_2013, varrassi_optical_2021}.
We also note that the introduction of the electron-hole interaction is necessary to obtain a good agreement for the onset of the experimental data,\cite{Exp_Kang} see Figure \ref{fig:ipa-mbse} b).

Having identified the key spectral peaks for both  phases, we now move to the analysis of how the corresponding excitons couple to the progressive ferroelectric distortion and evolve across the structural transition.

For the paraelectric phase, the dominant contribution to the $p_0$ peak arises from a triple-degenerate excitonic state with a BSE eigenvalue of \qty{3.393}{\electronvolt} (labelled $e_0$ in Figure \ref{fig:ipa-mbse}).
As shown by the fat-band analysis in Figure \ref{fig:fatband}, the associated BSE eigenvectors display comparable weight along the $\Gamma$–Z and $\Gamma$–X directions, involving transitions From O-$2p$ to Ti-$d$$t_{2g}$ states, reflecting the cubic symmetry of the phase.
Along these high-symmetry lines, the Ti-$t_{2g}$ manifold remains nearly non-dispersive, reflecting the localized character of these states; a similar behaviour has been reported for other cubic perovskites and has been related to the strong enhancement of the absorption peak due to excitonic effects.\cite{sponza_role_2013, varrassi_optical_2021}
\newline Figure \ref{fig:structures-spectra} highlights the changes in the isotropic absorption spectrum when the tetragonal distortion is gradually turned on. 
The previously identified $p_1$ and $p_2$ peaks can be directly traced back to the distortion-driven splitting of the $p_0$ peak, which first broadens and then separates into two distinct branches as the distortion grows (becoming fully resolved for distortions beyond the experimental value; see SM).
The splitting therefore introduces a distinct anisotropy in the dielectric response, with $p_1$ confined to the in-plane ($xx/yy$) channel and entirely absent from the out-of-plane ($zz$) spectrum.
\newline To clarify the microscopic origin of this anisotropy, we now examine the excitonic states associated with the split branches.
The contribution to the lower energy branch $p_1$ is dominated by the doubly degenerate exciton $e_1$
shown in Figure \ref{fig:fatband} for the ferroelectric structure ($c/a=1.069$) associated with a BSE eigenvalue of (\qty{3.96}{\electronvolt}).
Other excitons lying close in energy also provide non-negligible oscillator strength, but their fat-band distributions are largely similar to that of 
$e_1$. All of these states are doubly degenerate, consistent with  them leading to a peak in the $xx/yy$ degenerate component of the spectrum.
The $e_1$ excitonic wavefunction possesses spectral weights almost entirely confined along the $\Gamma$−Z direction, primarily involving O-$p_{x,y}$$\rightarrow$Ti-$d_{xy}$ transition. 
Because these transitions carry dipole matrix elements predominantly in the in-plane direction, $e_1$ contributes primarily to the $xx/yy$ component of the dielectric tensor, in line with the anisotropy observed for the $p_1$ peak.
\newline The $p_2$ feature in the ferroelectric structure is related to two excitonic states, $e_{2.1}$ and $e_{2.2}$ with eigenvalues at \qty{4.276}{\electronvolt}.
Both excitons receive their predominant contributions from out-of-plane transitions from O-2$p_{z}$ to conduction Ti-$d_{xz}$/$d_{yz}$ states. 
$e_{2.1}$ is doubly degenerate and has spectral weights mainly localised along $\Gamma-X$, albeit with much lower oscillator strengths compared to $e_1$; $e_{2.2}$ is instead predominantly localised at $\Gamma$ and non-degenerate.
These transitions carry dipole components with significant out-of-plane orientation, accounting for the appearance of spectral weight in the $zz$ response around the energy of the $p_2$ feature.
Lastly, the increasing $p_1$–$p_2$ separation reflects the distortion-driven splitting of the Ti $t_{2g}$ manifold. 
As the distortion grows, the energy separation between the Ti-$d_{xy}$ and $d_{xz,yz}$ states increases. As discussed above, the lower-energy branch $p_1$ is dominated by in-plane transitions to the Ti-$d_{xy}$ states (excitonic state $e_1$), whereas the higher-energy feature $p_2$ is associated with excitonic states with predominant transitions to Ti-$d_{xz,yz}$ states ($e_{2.1}$ and $e_{2.2}$).
Consequently, the $p_1$–$p_2$ separation increases with distortion, and the magnitude of the polar distortion therefore directly controls the energy splitting between the two features.
\newline Overall, the picture is consistent with the peak observed in the cubic structure absorption spectrum ($p_0$) developing from a triple degenerate exciton that, under structural symmetry breaking, splits into doubly degenerate and non-degenerate excitation that give rise to two separate spectral features. As a consequence of this, the first absorption peak in the ferroelectric structures is not simply shifted compared to the one present in the paraelectric phase, but actually arises due to excitations of different nature.

\section{Conclusions}
We have carried out a comprehensive \emph{ab-initio} study of the electronic and optical properties of \ch{PbTiO3} by combining $G_0W_0$ quasiparticle corrections with the solution of the Bethe-Salpeter equation, in the mBSE approximation, on a converged $\bm{k}$-mesh. By interpolating between the cubic and tetragonal structures, we traced the continuous evolution of the electronic properties and optical spectrum as the ferroelectric distortion is progressively switched on. The indirect and direct band-gaps obtained for the ferroelectric phase of PTO are respectively \qty{3.20}{\electronvolt} and \qty{3.49}{\electronvolt}, falling within the previously reported ranges from GW and experimental studies. 
\newline Solving the BSE reveals, for the paraelectric structure, an absorption peak at \qty{3.93}{\electronvolt} that is strongly enhanced by excitonic effects and evolves into two distinct spectral features as the polar distortion increases.
By analysing the BSE eigenvectors in $\bm{k}$-space, we show that the cubic structure peak is related to a triple degenerate exciton derived from transitions to d-t$_{2g}$ states localised along $\Gamma-Z$ and $\Gamma-X$.
Symmetry breaking partially lifts the degeneracy, giving rise to a doubly-degenerate lower excitonic state that contributes exclusively to the in-plane ($xx/yy$) dielectric channel, and a higher-energy set of excitonic states associated with the second spectral feature. 
This behavior follows directly from the different transition channels in the excitonic wavefunctions: in-plane O–2$p_{x,y}\rightarrow$Ti–$d_{xy}$ transitions for the lower branch and out-of-plane O–2$p_{z}\rightarrow$Ti–$d_{xz}$/$d_{yz}$ transitions for the higher branch.
The polar distortion therefore introduces the in-plane/out-of-plane anisotropy and controls the energy separation between the two branches. 

Beyond improving the accuracy of the first principles calculations for the properties of paraelectric and ferroelectric lead titanate already reported in the literature, our work demonstrates that polar lattice distortions, tunable through external electric fields, provide a knob for tailoring the optical absorption spectrum of a ferroelectric perovskite, opening the door to novel applications in opto-electronical devices.

\begin{acknowledgments}
The computational results have been achieved using the Austrian Scientific Computing (ASC) infrastructure.
\end{acknowledgments}

\section*{AUTHOR DECLARATIONS}

\subsection*{Conflict of Interest}
The authors have no conflicts to disclose.

\subsection*{Author Contributions}
\noindent{\textbf{Pietro Pacchioni:}} Investigation (lead); Data curation (lead); Formal analysis (equal); Writing – review \& editing (equal).
{\textbf{Cesare Franchini:}} Conceptualization (lead); Resources (lead); Supervision (equal); Formal analysis (equal);  Methodology (supporting); Writing – review \& editing (equal).
{\textbf{Lorenzo Varrassi:}} Methodology (lead); Supervision (equal); Formal analysis (equal); Writing – review \& editing (equal).
\section*{DATA AVAILABILITY}
The data that support the findings of this study are available
within the article and its supplementary material.

\bibliography{references}

\pagebreak

\section*{Supplementary Materials}
\subsection*{Further distorted structures}
As mentioned in \Cref{sec:methods} of the main text, performing relaxation using PBE functionals leads to overestimating the tetragonal distortion of TMO perovskites, thus only the interpolated structures with tetragonal distortion up to the experimental value were analyzed in detail.
\begin{table*}[h]
        \centering
        \begin{tabular}{c c c c c c c c c}
            \toprule
            Distortion & $a$ [\unit{\angstrom}] & $c/a$ & $E_g^{ind}$ [\unit{\electronvolt}] & $E_g^{dir}$ [\unit{\electronvolt}] & $E_g^{opt}$ [\unit{\electronvolt}] & $E_{xb}$ [\unit{\electronvolt}] & $E_{xb}^{(nd)}$ [\unit{\electronvolt}] & $\varepsilon_\infty^{-1}$\\
            \midrule
            \textbf{0\%}& 3.96 & 1 & 3.14 & 3.15 & 3.056 & 0.095 & 0.083 & 0.131 \\
            40\% & 3.95 & 1.027 & 3.14 & 3.19 & 3.097 & 0.098 & 0.086 & 0.132 \\
            60\% & 3.94 & 1.041 & 3.15 & 3.26 & 3.174 & 0.102 & 0.088 & 0.134 \\
            80\% & 3.93 & 1.055 & 3.18 & 3.36 & 3.280 & 0.107 & 0.090 & 0.136 \\
            \textbf{100\%} & 3.93 & 1.069 & 3.20 & 3.49 & 3.403 & 0.118 & 0.092 & 0.139 \\
            120\% & 3.92 & 1.083 & 3.23 & 3.63 & 3.543 & 0.132 & 0.094 & 0.141 \\
            160\% & 3.90 & 1.111 & 3.30 & 3.97 & 3.836 & 0.144 & 0.102 & 0.147 \\
            240\% & 3.87 & 1.167 & 3.49 & 4.25 & 4.223 & 0.172 & 0.153 & 0.161 \\
            \textbf{320\%} & 3.84 & 1.225 & 3.75 & 4.17 & 4.448 & 0.271 & 0.271 & 0.177 \\
            \bottomrule
        \end{tabular}
        \caption{Lattice constant $a$, tetragonal distortion ratio $c/a$, indirect band gap $E_g^{\mathrm{ind}}$, direct band gap $E_g^{\mathrm{dir}}$, optical gap $E_g^{\mathrm{opt}}$, exciton binding energy $E_{xb}$, first bright (non-dark) exciton binding energy $E_{xb}^{(\mathrm{nd})}$, and inverse dielectric screening $\varepsilon_\infty^{-1}$ for the interpolated structures, as obtained from $\mathrm{G_0W_0}$ and BSE calculations, including the overly distorted structures. The 320\% data corresponds to the relaxed structured obtained with PBE functionals.}
        \label{tab:electronic-properties1}
\end{table*}

\Cref{tab:electronic-properties1} extends \Cref{tab:electronic-properties} of the main text with further ferroelectric structures up to a distortion of 320\%, where 100\% corresponds to the experimental structure. The trends discussed in \Cref{sec:absorption-spectra} continue even for these unphysical structures, except for the direct band-gap $E_g^{dir}$ which abruptly decreases for the most elongated, which also leads to the breakdown of the scissor approximation in this extreme case.

\begin{figure}[h]
    \centering
    \includegraphics[width=\linewidth, trim={0 0 0 50pt}, clip]{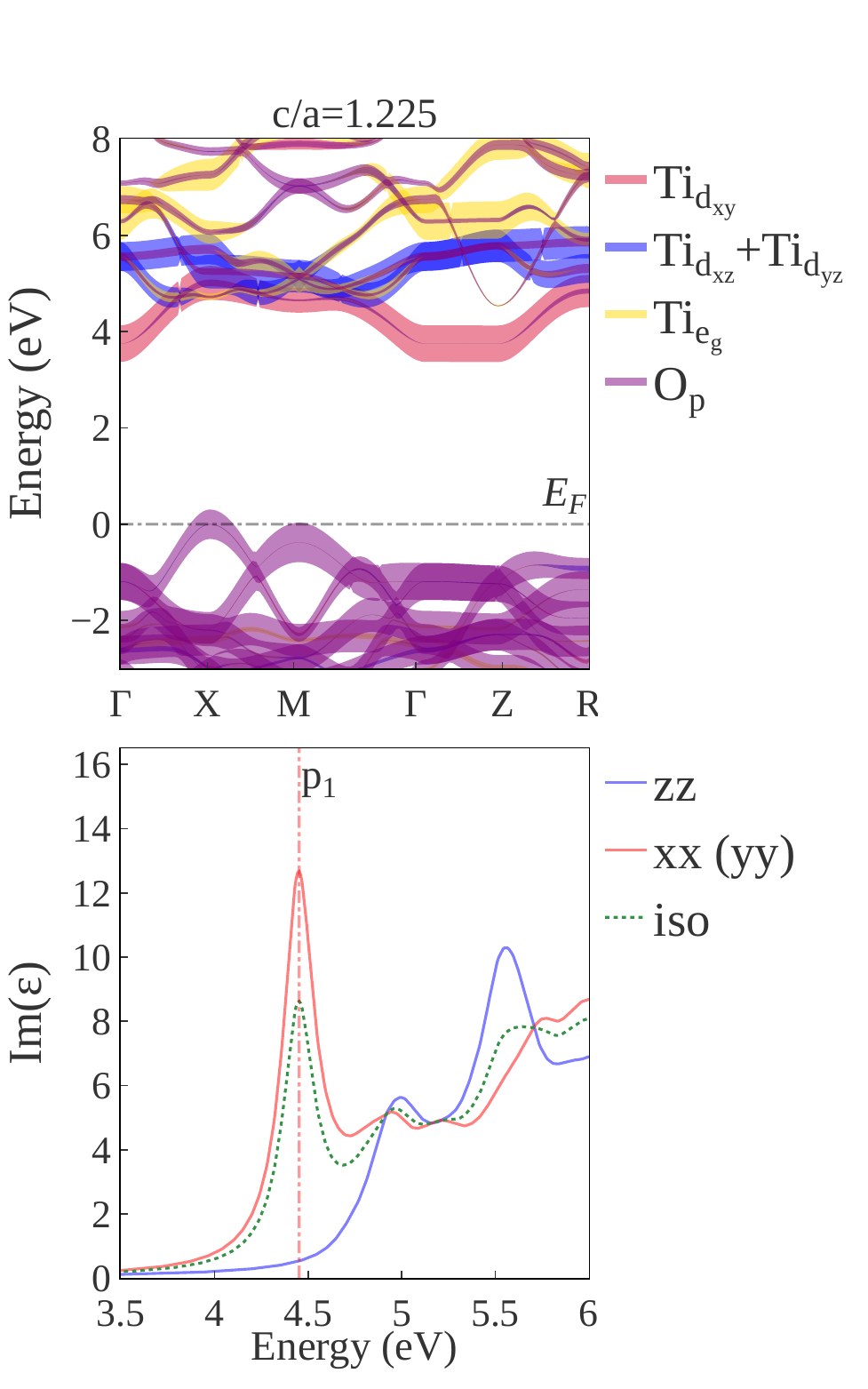}
    \caption{Equivalent of \Cref{fig:main-fig} for the 320\% distorted structure.}
    \label{fig:struc4}
\end{figure}

Nonetheless, significant differences can be found in the band structures and optical absorption spectra. For instance, \Cref{fig:struc4}, reports the projected band structure and imaginary  part  of  the  dielectric function for  the  most  distorted  structure. It can be noticed that the bands present significant more distortion then previously observed, and that significant mixing between the $e_g$ and $t_{2g}$ contributions arises. As expected from the rising optical band gap, the first absorption peak is further redshifted, but the second peak discussed in the main text is not discernible any-more, possibly due to the macroscopically changed and  more complicated electronic structure, or simply because it now resides outside the converged zone.

\subsection*{Scissor approximation accuracy}
Owing to the extreme computational cost of performing $G_0W_0$ calculations on a $k$-grid dense enough to obtain converged optical absorption spectra, the scissor approximation was employed instead. This consist in calculating the direct band gaps at both the PBE and $G_0W_0$  level with a less dense grid ($6\times6\times6$ in our case), finding the difference and then apply a corresponding rigid shift upwards to the valence bands obtained  on a denser grid with PBE. This approach is evidently only valid if the quasiparticle corrections do not significantly alter the shape itself of the bands, but only their energy.

\begin{figure*}
\centering
\begin{subfigure}[c]{0.49\textwidth}
    \includegraphics[width=\textwidth,trim={0pt, 0pt, 0pt, 0pt},clip]{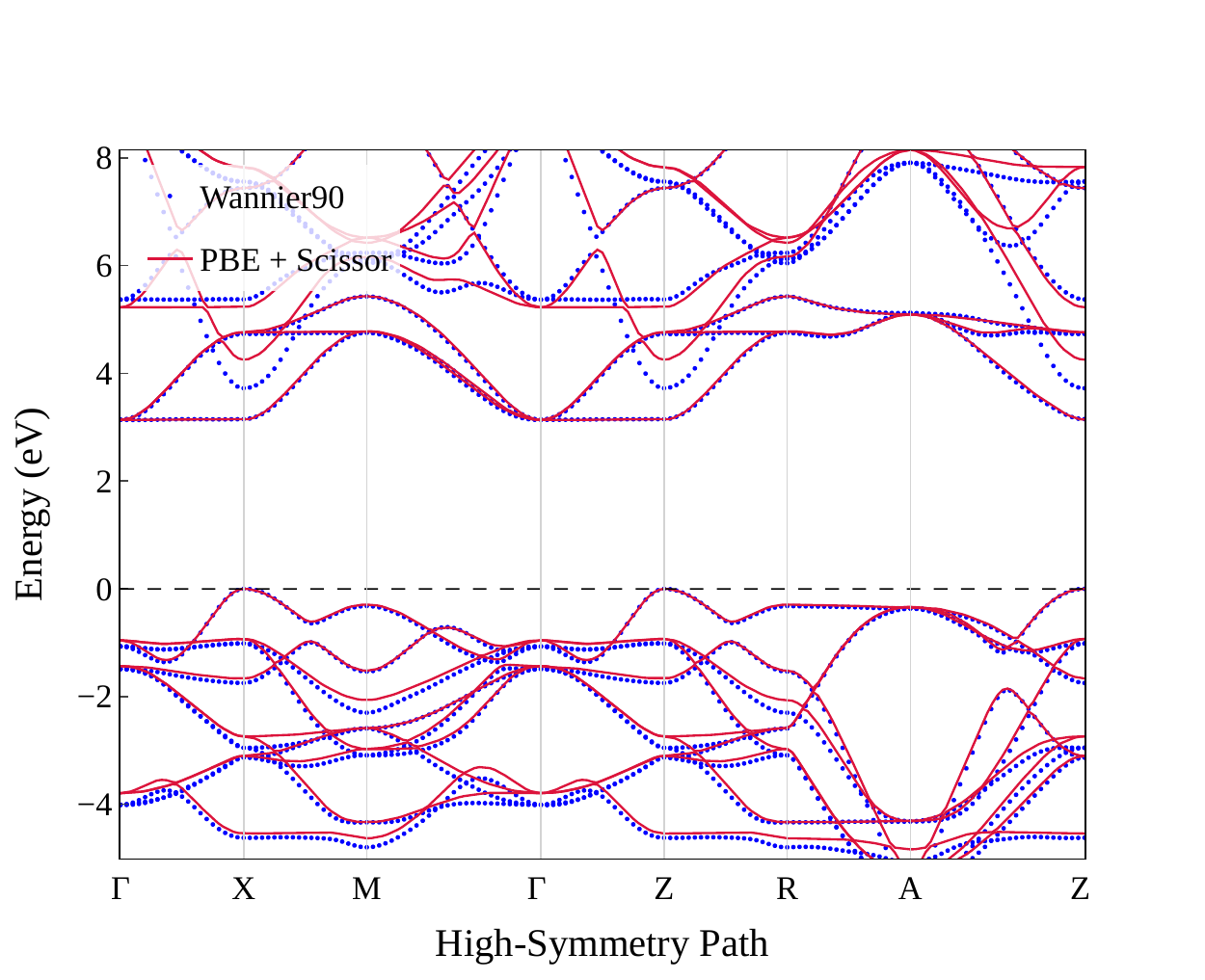}
    \caption{Cubic structure}
    \label{subfig:gw-dft0}
\end{subfigure}
\hfill
\begin{subfigure}[c]{0.49\textwidth}
    \includegraphics[width=\textwidth,trim={0pt, 0pt, 0pt, 0pt},clip]{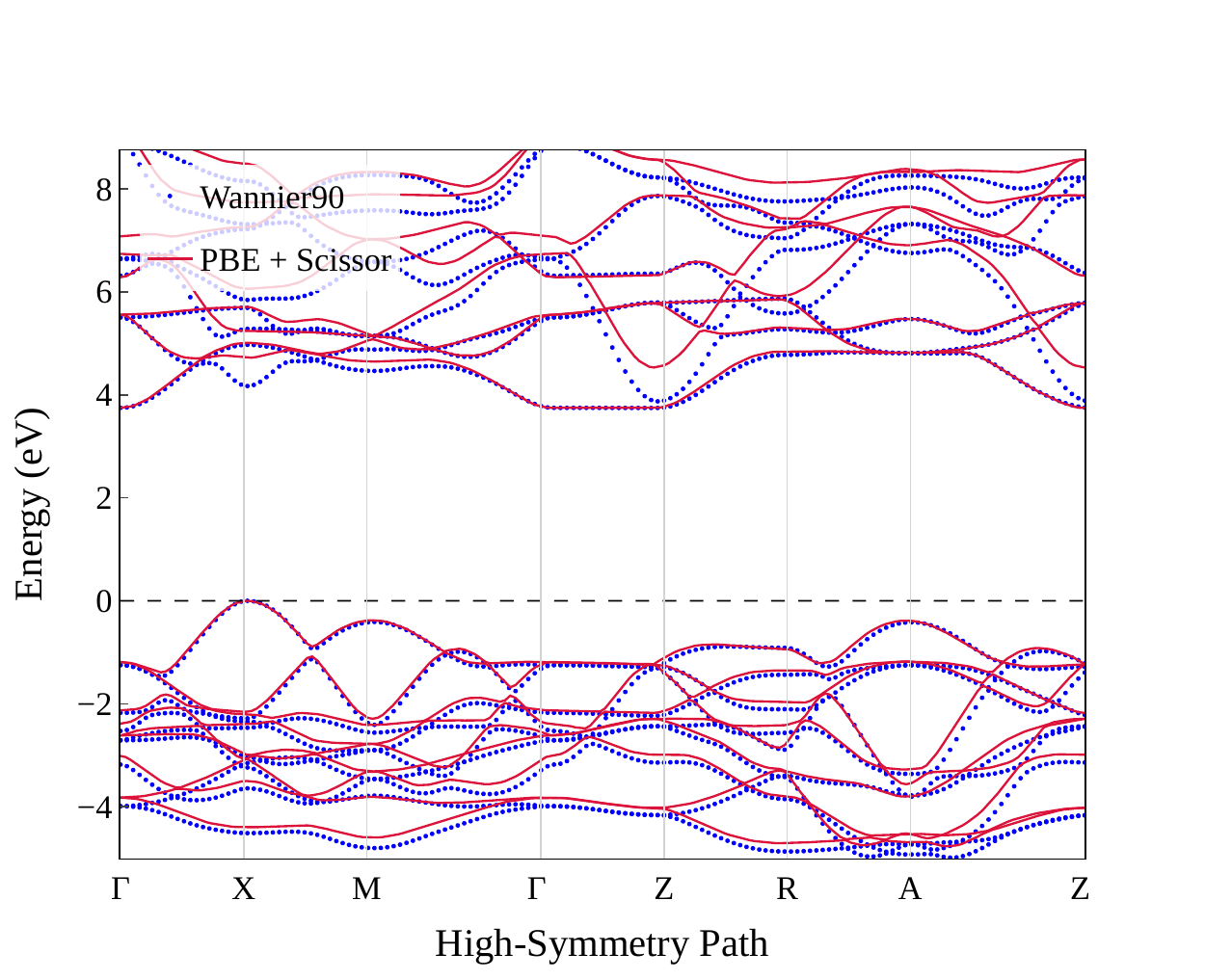}
    \caption{Overly distorted tetragonal structure}
    \label{subfig:gw-dft4}
\end{subfigure}
\caption{Comparison between the band structures of the 0\% and 320\% distorted PTO structures  calculated using Wannier interpolation on $G_0W_0$ data or by applying the scissor approximation to the PBE  results.}
\label{fig:gw-dft}
\end{figure*}

To verify this assumption, the full band structures of the extremal structures (0\% and 320\% distortion) were found starting from the $G_0W_0$ data using the Wannier interpolation method, as implemented by Wannier90~\cite{MOSTOFI2014}. Their comparison to properly shifted, PBE bands is presented in \Cref{fig:gw-dft}. The optically active bands of the cubic structure that are considered in the main text correspond almost perfectly, while a significant difference is present in the case  of  the overly  distorted  structure  near the X $k$-point, explaining the abrupt decrease in the direct band gap.

The scissor approximation is thus expected to produce the  correct optical properties for the cubic  and first distorted structures, but may result in inaccuracies when applied  to  the  unphysical cases discussed in the previous section.
\subsection*{model-BSE accuracy}
As discussed in the main text, the model-BSE approximation  replaces  the RPA  dielectric  function  with an analytical model, and it has been validated  on  other TMO  perovskites. However, its applicability is not  universal, and it is  therefore  important to verify  its  accuracy in the case of PTO.

Since we were not able to perform full BSE calculations using the parameters required  for convergence, we resorted to two other approximation methods already present in the literature: $k$-averaging and QP interpolation.

Given an $N\times N \times N$ target grid, the first approach requires first generating an $n\times n\times n$ base grid, made up of points at positions in reciprocal space $K_n$ with weights $W_n$, then performing BSE calculations on $m\times m\times m$ grids each centred at one $K_n$, where $m$ is such that $N=m\times n$. A weighted average is then performed over the dielectric functions. We applied this method only to the 320\% distorted structure, and we took $n=4$, $m=6$.

%\textcolor{red}{\textbf{write something about QP interpolation}}

\begin{figure*}
    \centering
    \includegraphics[width=\linewidth]{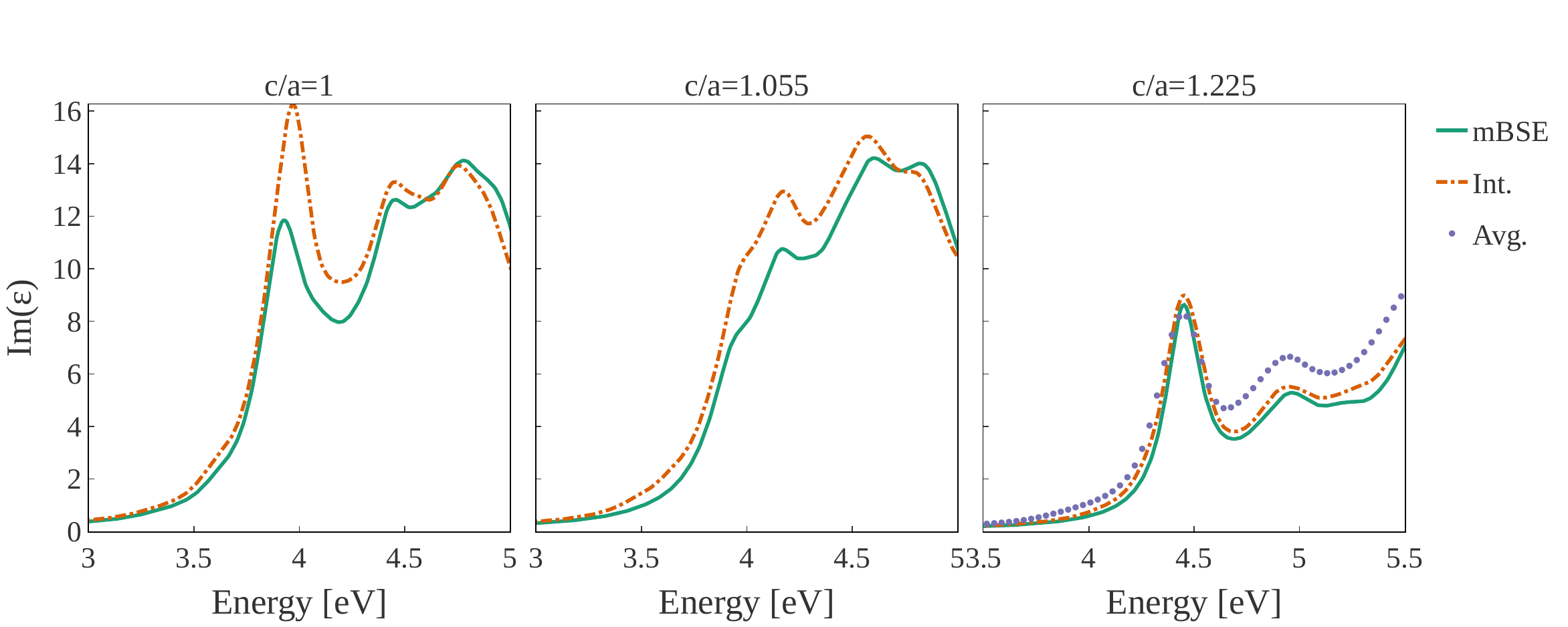}
    \caption{Averaged optical absorption spectra for three structures of different distortion obtained in the model-BSE approximation and through the QP interpolation method. The results of the averaging method on a $24\times24\times24$ $\bm{k}$-grid are also shown for the most distorted structure.   }
    \label{fig:mbse-int}
\end{figure*}

The absorption spectra resulting from the three approximations are compared in \Cref{fig:mbse-int}. Despite quantitative  differences  being present, especially for the cubic structure, the overall picture is consistent, as the position and shape of the peaks is constant,  suggesting  that the nature of the underlying excitations does not change. mBSE thus remains a valid, computationally approachable  method to study the optical properties of  PTO as its tetragonal distortion is varied.

\end{document}